\documentstyle[preprint,aps,epsfig]{revtex}

\tightenlines

\begin{document}

\title{Sigma-term physics in the perturbative chiral quark model}

\author{V. \ E. \ Lyubovitskij \footnotemark[1]\footnotemark[2], 
Th. \ Gutsche \footnotemark[1], Amand Faessler \footnotemark[1] and 
E. \ G. \ Drukarev \footnotemark[3] 
\vspace*{0.4\baselineskip}}
\address{
\footnotemark[1] 
Institut f\"ur Theoretische Physik, Universit\"at 
T\"ubingen, Auf der Morgenstelle 14, D-72076 T\"ubingen, Germany 
\vspace*{0.2\baselineskip}\\
\footnotemark[2]
Department of Physics, Tomsk State University, 634050 Tomsk, Russia 
\vspace*{0.3\baselineskip}\\
\footnotemark[3] 
Petersburg Nuclear Physics Institute, Gatchina, 
188350 St.Petersburg, Russia 
\vspace*{0.3\baselineskip}\\}

\maketitle

\vskip.5cm

\begin{abstract}
We apply the perturbative chiral quark model (PCQM) at one loop to analyse 
meson-baryon sigma-terms. Analytic expressions for these quantities are 
obtained in terms of fundamental parameters of low-energy pion-nucleon 
physics (weak pion decay constant, axial nucleon coupling, strong 
pion-nucleon form factor) and of only one model parameter (radius 
of the nucleonic three-quark core). Our result for the $\pi N$ sigma term 
$\sigma_{\pi N} \approx$ 45 MeV is in good agreement with the value deduced by 
Gasser, Leutwyler and Sainio using dispersion-relation techniques 
and exploiting the chiral symmetry constraints. 
\end{abstract}

\vspace*{\baselineskip}
\vskip1cm

       
\vskip.5cm


\section{Introduction} 

The meson-nucleon sigma-terms are fundamental parameters of low-energy hadron 
physics since they provide a direct measure of the 
scalar quark condensates in baryons and constitute a test for the mechanism of 
chiral symmetry breaking (for a review and recent advances in sigma-term 
physics see Refs.~\cite{Gasser_Ann,Reya,MENU99,DAPHNE99}). 
In particular, sigma-terms pose an important test for effective quark models 
in the low-energy hadron sector, since these quantities are mostly determined 
by the quark-antiquark sea and not by the valence quark contribution. 
The problem of the meson-nucleon sigma-terms is also closely related to 
properties of light hadron phenomenology such as the chiral expansion of 
baryon masses, the pseudoscalar meson-nucleon scattering lengths 
\cite{Gasser1,Meissner1,Leutwyler1}, properties of hadronic atoms \cite{Atom} 
and nuclear matter at finite temperature/density \cite{Faessler,Drukarev}. 

In current algebra the nucleon sigma-term $\sigma_{NN}^{ba}$ is derived 
from the low-energy theorem for meson-nucleon scattering. It is given by 
the double commutator of the axial-vector charge $Q_5^a$ 
with the chiral symmetry breaking ($\chi SB$) part of the Hamiltonian 
${\cal H}_{\chi SB}$, averaged over nucleon states $|N>$ \cite{Reya}:  
\begin{eqnarray}
\sigma_{NN}^{ab}\doteq <N|[Q^a_5 , \, [Q^b_5 , \, {\cal H}_{\chi SB}]]|N>. 
\end{eqnarray} 
In the QCD Hamiltonian the source of chiral symmetry breaking 
is simply the quark mass term with 
\begin{eqnarray}\label{S_i_QCD}
{\cal H}^{QCD}_{\chi SB} = \bar q {\cal M} q 
= m_u \, \bar u u + m_d \, \bar d d + m_s \, \bar s s 
= \sum_{i=u,d,s} m_i \, S_i^{QCD},   
\end{eqnarray}
where ${\cal M}={\rm diag}\{m_u, m_d, m_s \}$ is the mass matrix of current 
quarks and $S_i^{QCD}$ is the scalar density operator corresponding to the 
quark of i-th flavor. After straightforward use of current algebra one obtains
\begin{eqnarray}\label{Sigma-term}
\sigma_{NN}^{ab}=<N|\biggl\{\frac{\lambda^a}{2} , \,  
\biggl\{\frac{\lambda^b}{2} , \, {\cal M}\biggr\}\biggr\}|N>, 
\end{eqnarray}
where $\lambda^a$ $(a = 1, \ldots, 8)$ are the 
corresponding Gell-Mann flavor matrices and curly brackets refer to the 
anticommutator. In this paper we perform the calculations of the 
$\pi N$, $KN$ and $\eta N$ sigma-terms in the isospin symmetry limit with 
$m_u=m_d=\hat m$. With the use of key equation (\ref{Sigma-term}) the 
quantities of interest are given by 
\begin{eqnarray}\label{sigma_all}
& &\sigma_{\pi N} \doteq \sigma_{pp}^{11} = 
\hat m <p|\bar u u + \bar d d  |p>, \\
& &\sigma_{KN}^{u} \doteq \sigma_{pp}^{44} = 
\frac{\hat m + m_s}{2} <p|\bar u u + \bar s s|p>, \nonumber\\
& &\sigma_{KN}^{d} \doteq \sigma_{pp}^{66} = 
\frac{\hat m + m_s}{2} <p|\bar d d + \bar s s|p>, \nonumber\\
& &\sigma_{\eta N} \doteq \sigma_{pp}^{88} = 
\frac{1}{3} <p|\hat m (\bar u u + \bar d d)  + 2 m_s \bar s s  |p>.
\nonumber
\end{eqnarray}
where $|p>$ denotes a one-proton state. 
By analogy we can define the sigma-terms related to the $\Delta$-isobar by 
changing the proton state $|p>$ to the corresponding state $|\Delta>$ from 
the isomultiplet $\Delta(1230)$. 

The pion-nucleon sigma-term $\sigma_{\pi N}$ is equivalent to the value of 
the scalar nucleon form factor $\sigma(t)$ at zero value of momentum transfer 
squared: $\sigma_{\pi N} = \sigma(0)$ \cite{Leutwyler1}. On the other hand,   
the pion-nucleon $\sigma$ term is also related to the nucleon mass $m_N$ by 
means of the Feynman-Hellmann (FH) theorem \cite{Gasser_Ann,Feynman}:
\begin{eqnarray}\label{FHTh}
\sigma_{\pi N} = \hat m \frac{\partial m_N}{\partial \hat m}. 
\end{eqnarray} 
Both relations, $\sigma_{\pi N} = \sigma(0)$ and Eq. (\ref{FHTh}),   
are quite crucial as a consistent check of any approach 
applied to the study of the pion-nucleon $\sigma$-term. 

Other quantities which we consider are the strangeness content of 
the nucleon $y_N$ and the isovector $KN$ sigma-term $\sigma_{KN}^{I=1}$:  
\begin{eqnarray}
y_N=\frac{2<p|\bar s s|p>}{<p|\bar u u + \bar d d|p>}, \hspace*{.3cm}
\sigma_{KN}^{I=1}=\frac{\hat m + m_s}{4}  <p|\bar uu - \bar dd|p>,  
\end{eqnarray}
where the strange quark condensate $<p|\bar s s|p>$ is related to the nucleon 
mass by use of the FH theorem with 
$<p|\bar s s|p> = \partial m_N/\partial m_s$.    
The isovector $KN$ sigma-term is related to the flavor-asymmetric 
condensate $<p|\bar u u - \bar d d|p>$ of the proton. With the definitions of 
$y_N$ and $\sigma_{KN}^{I=1}$ we can relate $KN$ and $\eta N$ sigma-terms to 
the $\pi N$ sigma-term as 
\begin{eqnarray}
& &\sigma_{KN}^u=\sigma_{\pi N}(1+y_N)\frac{\hat m+m_s}{4\hat m}
+\sigma_{KN}^{I=1}, \hspace*{1cm} 
\sigma_{KN}^d=\sigma_{K N}^u - 2 \sigma_{KN}^{I=1}, \\
& &\sigma_{\eta N}=\sigma_{\pi N} \frac{\hat m + 2 y_N m_s}{3\hat m}. 
\nonumber
\end{eqnarray} 
In the literature other definitions of the KN sigma-terms can be 
found, namely \cite{VKM}-\cite{Borasoy},  
\begin{eqnarray}
\sigma_{KN}^{(1)} \equiv  \sigma_{KN}^{u} \,\,\,\,\, \mbox{and} \,\,\,\,\,  
\sigma_{KN}^{(2)}=\frac{\hat m +m_s}{2} <p|-\bar uu+2\bar dd+\bar ss|p>.
\end{eqnarray}
 
In the following we consider the relativistic quark model suggested in 
\cite{Gutsche} and recently extended in \cite{PCQM} for the study of 
low-energy properties of baryons, which are described as bound states of 
valence quarks surrounded by a cloud of Goldstone bosons ($\pi, K, \eta$) as 
required by chiral symmetry. Similar models have been studied in 
Refs. \cite{Theberge-Thomas,Oset,Chin}. We refer to our approach as the 
{\it perturbative chiral quark model} (PCQM). The PCQM is based on an  
effective chiral Lagrangian describing quarks as relativistic 
fermions moving in a self-consistent field (static potential). The latter 
is described by a scalar potential $S$ providing confinement of quarks 
and the time component of a vector potential $\gamma^0 V$ responsible for 
short-range fluctuations of the gluon field configurations \cite{Luscher}.  
Obviously, other possible Lorenz structures (e.g., pseudoscalar or 
axial) are excluded by symmetry principles. The model potential  
defines unperturbed wave functions of quarks which are subsequently used 
in the calculation of baryon properties. 
Interaction of quarks with Goldstone bosons is introduced on the basis 
of the nonlinear $\sigma$-model \cite{Gell-Mann_Levy}. When considering mesons 
fields as small fluctuations we restrict ourselves to the linear form of the 
meson-quark-antiquark interaction. With the derived interaction Lagrangian we 
do our perturbation theory in the expansion parameter $1/F$ 
(where $F$ is the pion leptonic decay constant in the chiral limit). We also 
treat the mass term of the current quarks as a small perturbation. 
Dressing the baryon three-quark core by the cloud of Goldstone mesons 
corresponds to inclusion of the sea-quark contribution. All calculations are 
performed at one loop or at order of accuracy 
$o(1/F^2, \hat m, m_s)$\footnote{
Throughout we use the Landau symbols $O(x)$ [$o(x)$] for quantities that
vanish like $x$ [faster than $x$] when $x$ tends to zero \cite{Atom}.}.  
The chiral limit with $\hat m, m_s \to 0$ is well defined. Electromagnetic 
gauge invariance is guaranteed at all steps of the calculations in a specific 
frame, that is the Breit frame (for details see \cite{PCQM}). 

In the present article we proceed as follows. First, we describe the basic 
notions of our approach: the underlying effective Lagrangian, the Dirac 
equation for quarks in the model potential, choice of free parameters and 
fulfilment of low-energy theorems, and finally, key perturbative equations 
for the calculation of baryon masses and sigma-terms. Other details 
(renormalization, gauge invariance and application to electromagnetic 
properties of nucleons) can be found in Ref. \cite{PCQM}. Next, we 
concentrate on the detailed analysis of the meson-nucleon sigma-terms 
in our approach: the chiral expansion of the $\pi N$ sigma-term in powers of 
pion mass and, after taking into account the kaon and $\eta$-meson loop 
contributions, model predictions for the $KN$, $\eta N$ sigma-terms and the 
strangeness content of the nucleon. Finally, we present our predictions for 
the $\Delta$-isobar sigma-terms. We compare all our predictions to results 
of other approaches, in particular, Chiral Perturbation Theory (ChPT), Heavy 
Baryon Chiral Perturbation Theory (HBChPT) and Lattice QCD. 

\section{Perturbative chiral quark model} 

We start with the zeroth-order Lagrangian describing relativistic quarks 
in an effective static potential $V_{eff}(r)=S(r)+\gamma^0 V(r)$ with 
$r=|\vec{x}|$ \cite{Gutsche,Oset,Fetter_Walecka}  
\begin{eqnarray}
{\cal L}^{(0)}(x)=\bar\psi(x)[i\not\!\partial -{\cal M}-S(r)-\gamma^0V(r)]
\psi(x) 
\end{eqnarray}
which obviously is not chiral invariant due to the presence of the quark mass 
term ${\cal M}$ and the scalar potential $S(r)$. To recover chiral invariance 
(of course, in the limit ${\cal M}\to 0$) we introduce the interaction of 
quarks with the octet of Goldstone pseudoscalar mesons $(\pi, K, \eta)$ by 
use of the nonlinear $\sigma$-model ansatz \cite{Gell-Mann_Levy}. We 
define the chiral fields using the {\it exponentional parametrization}    
$U=\exp[i\hat\Phi/F]$, where $F$ is the pion decay constant in the 
chiral limit. The octet matrix $\hat\Phi$ of pseudoscalar mesons is defined as
\begin{equation}
\frac{\hat\Phi}{\sqrt{2}}=
\sum_{i=1}^{8}\frac{\Phi_i\lambda_i}{\sqrt{2}}=
\left(
\begin{array}{ccc}
\pi^0/\sqrt{2} + \eta/\sqrt{6}\,\, & \,\, \pi^+ \,\, & \, K^+ \\
\pi^- \,\, & \,\, -\pi^0/\sqrt{2}+\eta/\sqrt{6}\,\, & \, K^0\\
K^-\,\, & \,\, \bar K^0 \,\, & \, -2\eta/\sqrt{6}\\
\end{array}
\right). 
\end{equation}
Kinetic term of meson fields ${\cal L}_{\Phi}$ and their interaction 
Lagrangian with quarks ${\cal L}_{int}$ are given by 
\begin{eqnarray}
{\cal L}_{\Phi} \,= \, \frac{F^2}{4} \, {\rm Tr} [ \, \partial_\mu U \, 
\partial^\mu U^\dagger \, ] \, = \, \frac{1}{2} \, ({\cal D}_\mu \Phi_i)^2 
\end{eqnarray}
and 
\begin{eqnarray}
{\cal L}_{int} \, = \, - \, \bar\psi(x) \, S(r) \, 
\biggl[ \frac{U \, + U^\dagger}{2} \, 
+ \, \gamma^5 \, \frac{U \, -  U^\dagger}{2} \biggr] \, \psi(x)
\, = \, - \, \bar\psi(x) \, S(r) \, \exp\biggl[ \, i \, \gamma^5 \, 
\frac{\hat\Phi}{F} \, \biggr] \, \psi(x)
\end{eqnarray}
where 
${\cal D}_\mu$ is the covariant chiral derivative \cite{Theberge-Thomas}, 
defined as 
$${\cal D}_\mu \Phi_i = \partial_\mu \Phi_i \, + \, 
\biggl(F \sin\frac{\Phi}{F} - \Phi \biggr) \,\, 
\partial_\mu \biggl(\frac{\Phi_i}{\Phi}\biggr) 
\,\,\,\,\,\,\, 
\mbox{with} \,\, \Phi=\sqrt{\Phi_i^2} $$   
The resulting Lagrangian ${\cal L}_{full}={\cal L}_{inv}+{\cal L}_{\chi SB}$ 
contains a chiral-invariant piece ${\cal L}_{inv}$ 
\begin{eqnarray}\label{Linv}
\hspace*{-1cm}
{\cal L}_{inv}(x)=\bar\psi(x)[i\not\! \partial -\gamma^0 V(r)]\psi(x)  
+ {\cal L}_{\Phi} + {\cal L}_{int} 
\end{eqnarray}
and includes a term ${\cal L}_{\chi SB}$ which explicitly breaks chiral 
symmetry:
\begin{eqnarray}\label{Lchi}
{\cal L}_{\chi SB}(x)=-\bar\psi(x){\cal M}\psi(x) 
-\frac{B}{8} {\rm Tr}\{ \hat\Phi , \, \{ \hat\Phi , \, {\cal M} \}\}  
\end{eqnarray}
corresponding to the mass terms for quarks and mesons. Here  
$B=-<0|\bar u u|0>/F^2$ is the low-energy constant which 
measures the vacuum expectation value of the scalar quark densities in the 
chiral limit \cite{Gasser_Leutwyler}. We rely on the standard picture of 
chiral symmetry breaking \cite{Gasser_Leutwyler} and for the masses 
of pseudoscalar mesons we use the leading term in their chiral expansion 
(i.e. linear in the current quark mass) 
\begin{eqnarray}\label{M_Masses}
M_{\pi}^2=2 \hat m B, \hspace*{.5cm} M_{K}^2=(\hat m + m_s) B, \hspace*{.5cm} 
M_{\eta}^2= \frac{2}{3} (\hat m + 2m_s) B. 
\end{eqnarray}
Meson masses obviously satisfy the Gell-Mann-Oakes-Renner and the 
Gell-Mann-Okubo relation $3 M_{\eta}^2 + M_{\pi}^2 = 4 M_{K}^2$. 
In the evaluation we use the following set of 
QCD parameters \cite{Gasser_Leutwyler_PR}:  
$\hat m$= 7 MeV, $m_s/\hat m$=25, $B=M_{\pi^+}^2/(2\hat m)$=1.4 GeV. 
For the pion decay constant $F$, defined in the chiral limit, we take the 
value of 88 MeV, calculated in ChPT \cite{Gasser1}. 

Our next approximation consists of a linearizing the resulting Lagrangian 
${\cal L}_{full}$ with respect to the chiral field $\hat\Phi$:  
\begin{eqnarray}\label{Linear}
U=\exp\biggl[i \, \frac{\hat\Phi}{F} \biggr] \simeq 
1 + i \frac{\hat\Phi}{F} + o\biggl(\frac{\hat\Phi}{F}\biggr).
\end{eqnarray}
In other words, we treat Goldstone fields as small fluctuations around 
the three-quark (3q) core. Considering the 3q baryon core as a zeroth-order 
approximation we take into account mesonic degrees of freedom by a 
{\it perturbative expansion}, i.e., expansion in terms of the coupling $1/F$. 
In the evaluation of matrix elements we restrict ourselves to the one-loop 
approximation, i.e. we perform our calculations to order of accuracy 
$o(1/F^2)$. Finally, with the use of ansatz (\ref{Linear}) our full 
Lagrangian ${\cal L}$ (see Eqs. (\ref{Linv}) and (\ref{Lchi})) reduces to 
the effective linearized Lagrangian 
\begin{eqnarray}\label{Lagrangian}
{\cal L}_{eff}(x)&=&\bar\psi(x)[i\not\!\partial-S(r)-\gamma^0 V(r)]\psi(x) 
\, + \, \frac{1}{2}(\partial_\mu \Phi_i)^2 \\
&-&  \bar\psi(x) S(r) i\gamma^5 \frac{\hat\Phi}{F} \psi(x) 
\, + \, {\cal L}_{\chi SB}(x).  
\nonumber
\end{eqnarray}

To describe the properties of baryons as bound states of quarks 
surrounded by a meson cloud we have to formulate perturbation theory. 
First, let us specify how we calculate the baryon mass 
spectrum (here and in the following we restrict considerations only to the  
nucleon and $\Delta$-isobar). In our approach the masses (energies) of 
the three-quark core for the nucleon and $\Delta$-isobar are degenerate and 
are related to the single quark energy 
${\cal E}_0$ by $E_0=m_N=m_\Delta=3\cdot {\cal E}_0$ (In this paper 
we do not discuss the removal of the spurious contribution to the baryon 
mass arising from the centre-of-mass motion of the bound state).   
The energy of the bound state $E_0$ satisfies the eigenequation 
$H_0|\phi_0>=E_0|\phi_0>$ with $H_0$ being the free quark Hamiltonian and 
$|\phi_0>$ being the unperturbed three-quark state. 
The single quark ground state energy ${\cal E}_0$ is obtained from the 
Dirac equation for the ground state quark wave function  (w.f.) 
$u_0(\vec{x})$ with 
\begin{eqnarray}\label{Dirac_Eq}
[-i\vec{\alpha}\vec{\nabla}+\beta S(r)+V(r)-{\cal E}_0]u_0(\vec{x})=0, 
\end{eqnarray}    
where we use the standard spinor algebra notations: 
$\vec{\alpha}=\gamma^0\vec{\gamma}$ and $\beta=\gamma^0$. 
The quark w.f. $u_0(\vec{x})$ belongs to the basis of potential 
eigenstates (including excited quark and antiquark solutions) used for 
expansion of the quark field operator $\psi(x)$. Here we restrict the  
expansion to the ground state contribution with 
$\psi(x)=b_0 u_0(\vec{x}) \exp(-i{\cal E}_0t),$ where $b_0$ is the 
corresponding single quark annihilation operator. 
In Eq. (\ref{Dirac_Eq}) we drop the current quark mass to simplify 
our calculational technique. Instead we consider the quark mass term as a 
small perturbation \cite{Gasser_Leutwyler_PR}. Inclusion of a finite current 
quark mass leads to a displacement of the single quark energy 
which, by Eq. (\ref{FHTh}), is relevant for the calculation of the sigma-term; 
a quantity which vanishes in the chiral limit. On the other hand, the effect 
of a finite current mass on observables which survive in the chiral limit, 
like magnetic moments, charge radii, etc., are quite negligible. 

In general, for a given form of the potentials $S(r)$ and $V(r)$ the Dirac 
equation (\ref{Dirac_Eq}) can be solved numerically. Here, for the sake of 
simplicity, we use a variational {\it Gaussian ansatz} \cite{Duck} for the 
quark wave function given by the analytical form: 
\begin{eqnarray}\label{Gaussian_Ansatz} 
u_0(\vec{x}) \, = \, N \, \exp\biggl[-\frac{\vec{x}^2}{2R^2}\biggr] \, 
\left(
\begin{array}{c}
1\\
i \rho \, \vec{\sigma}\vec{x}/R\\
\end{array}
\right) 
\, \chi_s \, \chi_f \, \chi_c
\end{eqnarray}      
where $N=[\pi^{3/2} R^3 (1+3\rho^2/2)]^{-1/2}$ is a constant fixed by the 
normalization condition $\int d^3x \, u^\dagger_0(x) \, u_0(x) \equiv 1$; 
$\chi_s$, $\chi_f$, $\chi_c$ are the spin, flavor and color quark wave 
functions, respectively. Our Gaussian ansatz contains two model parameters: 
the dimensional parameter $R$ and the dimensionless parameter $\rho$. 
The parameter $\rho$ can be related to the axial coupling constant $g_A$ 
calculated in zeroth-order (or 3q-core) approximation: 
\begin{eqnarray}
g_A=\frac{5}{3} \biggl(1 - \frac{2\rho^2} {1+\frac{3}{2} \rho^2} \biggr). 
\end{eqnarray}
For $\rho=0$ we obtain the nonrelativistic value of $g_A=5/3$. The parameter 
$R$ can be physically understood as the mean radius of the three-quark core 
and  is related to the charge radius $<r^2_E>^P$ of the proton in the 
zeroth-order approximation as 
\begin{eqnarray}
<r^2_E>^P_{3q-core}= \int d^3 x \, u^\dagger_0 (\vec{x}) \, \vec{x}^2 \, 
u_0(\vec{x}) \, = \, \frac{3R^2}{2} \, \frac{1 \, + \, \frac{5}{2} \, \rho^2}
{1 \, + \, \frac{3}{2} \, \rho^2}. 
\end{eqnarray}
In our calculations we use the value $g_A$=1.25 as obtained in 
ChPT \cite{Gasser1}. Therefore, we have only one free parameter $R$. 
In the numerical evaluation $R$ is varied in the region from 0.55 Fm  
to 0.65 Fm corresponding to a change of $<r^2_E>^P_{3q-core}$ in the 
region from 0.5 Fm$^2$ to 0.7 Fm$^2$. The use of the Gaussian ansatz 
(\ref{Gaussian_Ansatz}) in its exact form restricts the scalar confinement 
potential $S(r)$ to  
\begin{eqnarray}
S(r) = \underbrace{\frac{1 \, - \, 3\rho^2}{2 \, \rho R}}_{= \, M} 
\,\, + \,\,
\underbrace{\frac{\rho}{2R^3}}_{= \, c} \, r^2 = M \,\, + \,\, c \, r^2,   
\end{eqnarray}
expressed in terms of the parameters $R$ and $\rho$. 
The constant part of the scalar potential $M$ can be interpreted 
as the constituent mass of the quark, which is simply a displacement of the 
current quark mass due to the potential $S(r)$. The parameter $c$ is the 
coupling defining the radial (quadratic) dependence of the scalar potential. 
Numerically, for our set of parameters, we get $M=230\pm 20$ MeV 
and $c=0.08\pm 0.01$ GeV$^3$. These and the following error bars 
in our results correspond to the variation of the parameter $R$. For the 
vector potential we get the following expression 
\begin{eqnarray}
V(r) = {\cal E}_0 \,\,  - \,\, \frac{1 \, + \, 3\rho^2}{2 \, \rho R}
\,\, + \,\, \frac{\rho}{2R^3} \, r^2, 
\end{eqnarray}
where the single quark energy ${\cal E}_0$ is a free parameter in the 
Gaussian ansatz. For our purposes (calculation of the sigma-terms) the 
magnitude of the quark energy in zeroth-order ${\cal E}_0$ does not influence 
the result. Nevertheless, in view of Eq. (\ref{FHTh}), the shift of 
${\cal E}_0$ due to a finite quark mass $m$ will contribute to the 
meson-baryon sigma-terms. The displacement of the quark energy in an 
expansion to first order in $m$, with the use our Gaussian ansatz,  
is 
\begin{eqnarray}
{\cal E}_0\to {\cal E}_0(m)={\cal E}_0 + \gamma m + o(m), 
\end{eqnarray}
where $\gamma$ 
is the relativistic reduction factor  
\begin{eqnarray}
\gamma=\frac{1-\frac{3}{2}\rho^2}{1+\frac{3}{2}\rho^2}=
\frac{9}{10}g_A-\frac{1}{2}.
\end{eqnarray}
In the nonrelativistic limit $(\rho = 0)$ we get $\gamma=1$ and, therefore, 
the corresponding displacement is equal to the current quark mass. In the 
relativistic picture, for $\rho^2 > 0$,  we definitely have  
$\gamma < 1$. Particularly for our choice of $g_A=1.25$ we get $\gamma=5/8$.  

In our approach the PCAC requirement and consequently the Goldberger-Treiman 
relation are fulfiled. In other words, the expectation value of the 
pseudoscalar isovector density 
$J_i \, = \, (1/F) \, S(r)\,  \bar\psi \, i \, \gamma_5 \, \tau_i \, \psi$ 
between unperturbed 3q states $|\phi_0>$ states leads to the pion-nucleon 
constant $G_{\pi NN}$ and we arrive at the Goldberger-Treiman relation 
between the couplings $G_{\pi NN}$ and $g_A$ with 
\begin{eqnarray}
G_{\pi NN}=\frac{5m_N}{3F} \, \biggl(1 \, - \, \frac{2\rho^2}
{1 \, + \, \frac{3}{2} \, \rho^2}\biggr) 
\equiv \frac{m_N}{F} \, g_A .  
\end{eqnarray}
The same condition holds even for their form factors. The analytical 
expression for the pion-nucleon form factor in the chiral limit is given by 
\begin{eqnarray}
G_{\pi NN}(Q^2)=\frac{m_N}{F} \, g_A(Q^2)= \frac{m_N}{F} \, g_A \, 
F_{\pi NN}(Q^2) 
\end{eqnarray}
where $Q^2$ is the squared Euclidean momentum of the pion and 
$F_{\pi NN}(Q^2)$ is the $\pi NN$ form factor normalized to unity at zero 
recoil $Q^2=0$: 
\begin{eqnarray}
F_{\pi NN}(Q^2) = \exp\biggl(-\frac{Q^2R^2}{4}\biggr) \biggl\{ 1 \, + \, 
\frac{Q^2R^2}{8} \biggl(1 \, - \, \frac{5}{3g_A}\biggr)\biggr\} .  
\end{eqnarray}

Following the Gell-Mann and Low theorem \cite{Gell-Mann_Low} the energy shift 
$\Delta E_0$ of the three-quark ground state due to the interaction with 
Goldstone mesons is given by the expression 
\begin{eqnarray}\label{Energy_shift}
\hspace*{-.8cm} 
\Delta E_0=<\phi_0| \, \sum\limits_{i=1}^{\infty} \frac{(-i)^n}{n!} \, 
\int \, i\delta(t_1) \, d^4x_1 \ldots d^4x_n \, 
T[{\cal H}_I(x_1) \ldots {\cal H}_I(x_n)] \, |\phi_0>_{c} 
\end{eqnarray}
where 
\begin{eqnarray}
{\cal H}_I(x)= \bar\psi(x) i\gamma^5 \frac{\hat\Phi(x)}{F} S(r)\psi(x)
\end{eqnarray} 
is the interaction Hamiltonian and subscript "c" refers to contributions 
from connected graphs only. We evaluate Eq. (\ref{Energy_shift}) at one loop 
with $o(1/F^2)$ using Wick's theorem and the appropriate propagators. 
For the quark field we use a Feynman propagator for a fermion in a binding 
potential. By restricting the summation over intermediate quark states 
to the ground state we get   
\begin{eqnarray} 
\hspace*{-.75cm} 
iG_\psi(x,y)=<\phi_0|T\{\psi(x)\bar\psi(y)\}|\phi_0> \to 
u_0(\vec{x}) \bar u_0(\vec{y})\exp[-i{\cal E}_0(x_0-y_0)]\theta(x_0-y_0). 
\end{eqnarray} 
For meson fields we use the free Feynman propagator for a boson field with  
\begin{eqnarray}
i\Delta_{ij}(x-y)=<0|T\{\Phi_i(x)\Phi_j(y)\}|0>=\delta_{ij} 
\int\frac{d^4k}{(2\pi)^4i}\frac{\exp[-ik(x-y)]}{M_\Phi^2-k^2-i\epsilon}.  
\end{eqnarray}

\section{Meson-baryon sigma-terms in the PCQM} 

The scalar density operators $S_i^{PCQM}$ $(i=u, d, s)$, relevant for the 
calculation of the meson-baryon sigma-terms in the PCQM, are defined as the  
partial derivatives of the model $\chi SB$ Hamiltonian 
${\cal H}_{\chi SB}=-{\cal L}_{\chi SB}$ with respect to the current 
quark mass of i-th flavor $m_i$. Note that the nondiagonal term 
in ${\cal H}_{\chi SB}$ which is proportional to $(m_u-m_d) \pi^0 \eta$ 
vanishes because we apply the isospin limit with $m_u=m_d$.  Here we obtain 
\begin{eqnarray}
S_i^{PCQM} \doteq \frac{\partial {\cal H}_{\chi SB}}{\partial m_i} 
\, = \, S_i^{val} \, + \, S_i^{sea}, 
\end{eqnarray}
where $S_i^{val}$ is the set of valence-quark operators 
coinciding with the ones obtained from the QCD Hamiltonian (\ref{S_i_QCD}) 
\begin{eqnarray}\label{S_val}
S_u^{val} = \bar u u, \hspace*{1cm}  
S_d^{val} = \bar d d, \hspace*{1cm}  
S_s^{val} = \bar s s .  
\end{eqnarray}
The set of sea-quark operators $S_i^{sea}$ arises from the pseudoscalar 
meson mass term (due to Eq. (\ref{M_Masses})) with  
\begin{eqnarray}\label{S_sea}
S_u^{sea} &=& B \, \biggl\{ \, \pi^+ \, \pi^- \, + \, 
\frac{\pi^0 \, \pi^0}{2}  \, + \, K^+ \, K^- \, + \, 
\frac{\eta^2}{6}  \, \biggr\}, \\
S_d^{sea} &=& B \, \biggl\{ \, \pi^+ \, \pi^- \, + \, 
\frac{\pi^0 \, \pi^0}{2}  \, + \, K^0 \, \bar K^0 \, + \, 
\frac{\eta^2}{6}  \, \biggr\}, \nonumber\\
S_s^{sea} &=& B \, \biggl\{ \, K^+ \, K^- \, + \, K^0 \, \bar K^0 \, + \, 
\frac{2}{3} \, \eta^2  \, \biggr\}. \nonumber 
\end{eqnarray} 
To calculate meson-baryon sigma-terms, equivalent to the definition of 
Eq. (\ref{sigma_all}), we perform the perturbative expansion 
for the matrix element of the scalar density operator $S_i^{PCQM}$ 
between unperturbed 3q-core states:  
\begin{eqnarray}\label{Key_Eq} 
\hspace*{-.75cm}
<\phi_0| \, \sum\limits_{n=0}^{\infty} \frac{(-i)^n}{n!}
\int  \, d^4x_1 \ldots \int  d^4x_n \, \, 
T[{\cal H}_I(x_1) \ldots {\cal H}_I(x_n) \,  S_i^{PCQM} ] \, |\phi_0>_c . 
\end{eqnarray}
In the evaluation of Eqs. (\ref{Energy_shift}) and (\ref{Key_Eq}) 
we project the matrix elements on the respective baryon states. 
Baryon wave functions for nucleon and delta states are conventionally set up 
by the product of the $SU(6)$ spin-flavor w.f. and $SU(3)_c$ color w.f. 
(see details in \cite{Close}), where the nonrelativistic single quark 
spin w.f. is replaced by the relativistic ground state solution of 
Eq. (\ref{Gaussian_Ansatz}). 

The diagrams that contribute to the energy shift $\Delta E_0$ at one loop 
are shown in Fig.1: meson cloud (Fig.1a) and meson exchange diagram 
(Fig.1b). The explicit expressions for the nucleon and $\Delta$-isobar masses 
including one-loop corrections are given by 
\begin{eqnarray}\label{Baryon_mass}
&&m_B=3 \, ({\cal E}_0 + \gamma\hat{m}) \, + \, 
d_B^\pi \, \Pi (M_\pi^2) \, + \, d_B^K \, \Pi (M_K^2) 
\, + \, d_B^\eta \, \Pi (M_\eta^2) \\
\mbox{with} \,\,\,\,\, & &d_N^\pi=\frac{171}{400}, \,\,\,   
d_\Delta^\pi=\frac{11}{19}d_N^\pi, \,\,\, d_B^K=\frac{6}{19}d_N^\pi, \,\,\, 
d_N^\eta=\frac{1}{57}d_N^\pi, \,\,\, d_\Delta^\eta=\frac{5}{57}d_N^\pi , 
\nonumber
\end{eqnarray} 
where $d_B^\Phi$ with $\Phi=\pi, K$ or $\eta$ and $B=N$ or $\Delta$ are the 
recoupling coefficients defining the partial contribution of the $\pi$, $K$ 
and $\eta$-meson cloud to the energy shift of the nucleon and 
$\Delta$-isobar, respectively. The self-energy operators $\Pi (M_\Phi^2)$, 
corresponding to meson cloud contributions with definite flavor, differ only 
in their value for the meson mass and are given by 
\begin{eqnarray}\label{Sigma_Phi}
\Pi (M_\Phi^2) \, = \, - \, \biggl(\frac{g_A}{\pi F}\biggr)^2 \,\, 
\int\limits_0^\infty \frac{dp \, p^4}{p^2+M_\Phi^2} \,\, F_{\pi NN}^2(p^2) .  
\end{eqnarray}
where $p=|\vec{p}|$ is the absolute value of the three momentum of the meson. 
Using Eqs. (\ref{Baryon_mass}) and (\ref{Sigma_Phi}) we obtain the 
expressions for the baryon (nucleon and $\Delta$-isobar) masses
\begin{eqnarray}\label{Baryon_mass_exp}
m_B &=&  \underbrace{3 {\cal E}_0 + \Pi(0) \cdot 
\sum\limits_{\Phi= \pi, K, \eta} d_B^\Phi }_{=\stackrel{0}m_B} \,\, + \,\,  
\underbrace{3 \gamma\hat{m} + \sum\limits_{\Phi=\pi, K, \eta} 
d_B^\Phi \cdot [\Pi (M_\Phi^2) - \Pi (0) ]}_{=\Delta m_B} \\
&=& \stackrel{0}m_B \, + \, \Delta m_B . \nonumber
\end{eqnarray} 
The baryon mass in the chiral limit $(\hat m, m_s \to 0)$ is 
$\stackrel{0}m_B$. Meson loops also contribute to $\stackrel{0}m_B$, since  
in the chiral limit quarks interact with massless mesons. In the calculations 
of the one-loop meson contributions to the baryon masses we do not take into 
account the modification of the quark w.f. due to a small but finite current 
quark mass. This effect is strongly suppressed and therefore, as mentioned 
above,  we work at the order of accuracy $o(\hat m, m_s, 1/F^2)$. 

At the same level of accuracy $o(\hat m, m_s, 1/F^2)$ we calculate the 
meson-baryon sigma-terms using Eq. (\ref{Key_Eq}). The following diagrams 
contribute to the sigma-terms up to the one-loop level: tree level diagram 
(Fig.2a) with the insertion of the valence-quark scalar density $S_i^{val}$ 
into the quark line, meson cloud (Fig.2b) and meson exchange diagrams (Fig.2c) 
with insertion of the sea-quark scalar density $S_i^{sea}$ to the meson 
line. We neglect diagrams with an insertion of $S_i^{val}$ to the quark lines 
in the second-order graphs (like in Fig.1a and Fig.1b). These terms  
are proportional to $\hat m/F^2$ and therefore are of higher order. 
First, we consider the $\pi N$ sigma-term. The analytical 
expression for this quantity is derived as 
\begin{eqnarray}\label{Sigma_piN}
\sigma_{\pi N} = \hat m \, <p|S_u^{PCQM} \, + S_d^{PCQM}|p>  
= 3 \gamma\hat{m} \, + \, \sum\limits_{\Phi=\pi, K, \eta} 
d_N^\Phi \cdot \Gamma (M_\Phi^2)  
\end{eqnarray} 
where the first term of the right-hand side of Eq. (\ref{Sigma_piN}) 
corresponds to the valence quark, the second to the sea quark contribution. 
The vertex function $\Gamma (M_\Phi^2)$, as derived by use of Eq. 
(\ref{Key_Eq}), can be related to the partial derivative of the self-energy 
operator $\Pi (M_\Phi^2)$ with respect to the nonstrange current quark mass 
$\hat m$ with  
\begin{eqnarray}\label{Gamma_piN}
\Gamma (M_\Phi^2) \, = \, 
\hat{m} \, \frac{\partial}{\partial \hat{m}} \, \Pi (M_\Phi^2) . 
\end{eqnarray}   
The derivative $\hat{m}\partial/\partial{\hat{m}}$ is equivalent to the one  
with respect to the meson masses \cite{Gasser_Ann}:
\begin{eqnarray}
\hat{m}\frac{\partial}{\partial\hat{m}} \Pi (M_\Phi^2) = 
M_\pi^2\biggl(\frac{\partial}{\partial M_\pi^2} + 
\frac{1}{2}\frac{\partial}{\partial M_K^2} + 
\frac{1}{3}\frac{\partial}{\partial M_\eta^2} \biggr) \Pi (M_\Phi^2).  
\end{eqnarray}
Using Eqs. (\ref{Baryon_mass_exp}), (\ref{Sigma_piN}) and (\ref{Gamma_piN}) 
we directly prove the Feynman-Hellmann theorem of Eq. (\ref{FHTh}); hence our 
approach is consistent to order of accuracy $o(\hat m, m_s, 1/F^2)$. 
Our next point is the analysis of the chiral expansion for the $\pi N$ 
sigma-term $\sigma_{\pi N}^\pi$ when we restrict to the 
two-flavor picture, i.e., we take into account only pion contributions. The 
expansion for $\sigma_{\pi N}^\pi$ in the PCQM is given by an analytic 
expression in terms of fundamental constants (the couplings $g_A$ and $F$, 
the quark and pion masses $\hat m$ and $M_\pi$)  
\begin{eqnarray}\label{Sigma_piN_exp}
\hspace*{-1cm}
\sigma_{\pi N}^\pi = \frac{3\gamma}{2B} M_\pi^2 \, + 
\, \frac{d_N^\pi}{(2\pi)^{3/2}} \, 
\biggl(\frac{g_A}{F}\biggr)^2  \frac{M_\pi^2}{R}  
\sum\limits_{N=0}^{\infty}  \frac{(-M_\pi R)^N}{2^{N/2}} \, 
\frac{\Gamma(\frac{1}{2})}{\Gamma(\frac{N+1}{2})} 
\, (N+2) \, {\cal F}_N(\gamma)
\end{eqnarray} 
where 
$$ 
{\cal F}_N(\gamma) \, = \, 1 \, + \, 
\biggl(\frac{1-\gamma}{1+2\gamma}\biggr)\frac{N-1}{2} \, + \, 
\biggl(\frac{1-\gamma}{1+2\gamma}\biggr)^2 \frac{(N-1)(N-3)}{16} 
$$
is a polynomial in the relativistic factor $\gamma$ introduced before. 
Explicitly, up to order $o(M_\pi^5)$ we obtain:  
\begin{eqnarray}\label{Sigma_piN_exp4}
\sigma_{\pi N}^\pi = k_2 M_\pi^2 + 
\frac{3}{2} k_3 M_\pi^3 + 2 k_4 M_\pi^4 +\frac{5}{2} k_5 M_\pi^5 + o(M_\pi^5) 
\end{eqnarray} 
where $k_i$ are the expansion coefficients evaluated as 
\begin{eqnarray}\label{c2_c4}
& &k_2 \, = \, \frac{3\gamma}{2B}\, + \, \frac{171}{200}\, 
\biggl(\frac{g_A}{F}\biggr)^2 \, \frac{1 + 2\xi - 3\xi^2}{(2\pi)^{3/2}R}, 
\hspace*{.5cm}
k_3 \, = \, - \, \frac{171}{800\pi} \, \biggl(\frac{g_A}{F}\biggr)^2 \\
& &k_4 \, = \, \frac{171}{200} \, \biggl(\frac{g_A}{F}\biggr)^2 \, R \,\,   
\frac{1 - 2\xi - \xi^2}{(2\pi)^{3/2}}, \hspace*{.5cm} 
k_5 \, = \, - \, \frac{171}{1600} \, 
\biggl(\frac{g_A}{F}\biggr)^2 \, \frac{R^2}{\pi} \,\, ( 1 - 4\xi ), \nonumber\\
& &\hspace*{.3cm}\xi=\frac{1-\gamma}{4(1+2\gamma)}.  
\nonumber
\end{eqnarray}
In the present approach the coefficient $k_3$ defines the leading 
nonanalytic contribution (LNAC) to $\sigma_{\pi N}$. Both, nucleon 
and $\Delta$-isobar degrees of freedom as intermediate states in the 
meson-loop diagrams contribute to the coefficient $k_3$. The nucleonic 
contribution to the coefficient $k_3$ is $k_3^N=-3g_A^2/(32\pi F^2)$. 
The $\Delta$-isobar also contributes to $k_3$ with 
$k_3^\Delta=32/25 \, k_3^N$. In total, our LNA coefficient $k_3$ results 
in $k_3=k_3^N+k_3^\Delta=-171g_A^2/(800\pi F^2)$ as also obtained in a 
similar quark model approach \cite{Glozman}. One should remark that our 
LNAC disagrees with a similar quantity given in effective hadronic 
approaches  
\cite{Leutwyler1,Gasser_Leutwyler_PR,Jenkins_Manohar,Henley_Thirring}  
and in the cloudy bag model \cite{Thomas_Krein}. In latter approaches 
the unperturbed nucleon and $\Delta$ states are not degenerate. Therefore, 
only the loop diagram with $\pi N$ as an intermediate state contributes to 
the LNAC which is equal to $-3g_A^2/(32\pi F^2)$ and coincides with our 
$k_3^N$. The $\pi\Delta$ loop diagram contributes to the next-to-leading 
order nonanalytic term. 

Numerically, the contribution of the valence quarks to 
the $\pi N$ sigma-term is $13.1$ MeV, the contribution of sea quarks at 
order $o(M_\pi^2)$ is $66.9 \pm 5.7$ MeV. Higher-order contributions of 
the sea quarks are $(3k_3/2) M_\pi^3=-57.2$ MeV, 
$2 k_4 M_\pi^4=27.7 \pm 2.3$ MeV,   
$(5k_5/2) M_\pi^5=-10\pm 1.7$ MeV and $o(M_\pi^5) \approx $ 2.8 MeV.  
Therefore, at order $o(M_\pi^5)$ in the two-flavor picture we have the 
following result for the $\pi N$ sigma-term $\sigma_{\pi N}^{\pi}$ 
(again, superscript $\pi$ refers to the SU(2) flavor picture) of 
\begin{eqnarray}
\sigma_{\pi N}^{\pi(5)} = 40.5 \pm 5 \, {\rm MeV},
\end{eqnarray}
compared to the full calculation of the $\pi N$ 
sigma-term of Eq. (\ref{Sigma_piN_exp}) of  
\begin{eqnarray}\label{Sigma_piN_pi}
\sigma_{\pi N}^{\pi} = 43.3 \pm 4.4 \,  {\rm MeV}.
\end{eqnarray}
As we mentioned before, the error bars are due to a variation of the range 
parameter $R$ of the quark wave function (\ref{Gaussian_Ansatz}) from  
0.55 Fm to 0.65 Fm. 

To complete our estimate of the $\pi N$ sigma-term we evaluate the additional 
contributions of kaon and $\eta$-meson loops, $\sigma_{\pi N}^{K}$ and 
$\sigma_{\pi N}^{\eta}$, where superscripts $K$ and $\eta$ refer to the 
respective meson cloud contribution. These terms are significantly 
suppressed relative to the pion cloud and to the valence quark contributions. 
For illustration, we give the expressions for $r_K$ (ratio of kaon to 
pion cloud contributions) and $r_\eta$ (ratio of $\eta$-meson to pion 
cloud contributions): 
\begin{eqnarray}
r_K &=& \frac{\sigma_{\pi N}^{K}}{\sigma_{\pi N}^{\pi} - 3\gamma \hat{m}} 
= \frac{d_N^K}{2d_N^\pi} \,\, \cdot \,\, 
\frac{\displaystyle{\int\limits_0^\infty\frac{dp\, p^4}{(p^2+M_K^2)^2}\,\, 
F_{\pi NN}^2(p^2)}}
{\displaystyle{\int\limits_0^\infty
\frac{\displaystyle{dp\, p^4}}{\displaystyle{(p^2+M_\pi^2)^2}}\,\, 
F_{\pi NN}^2(p^2)}} , 
\label{ratio_K_eta1}\\
& &\nonumber\\
r_\eta &=& \frac{\sigma_{\pi N}^{\eta}}
{\sigma_{\pi N}^{\pi} - 3\gamma \hat{m}} 
= \frac{d_N^\eta}{3d_N^\pi} \,\, \cdot \,\, 
\frac{\displaystyle{\int\limits_0^\infty\frac{dp\, p^4}{(p^2+M_\eta^2)^2}\,\, 
F_{\pi NN}^2(p^2)}}
{\displaystyle{\int\limits_0^\infty
\frac{\displaystyle{dp\, p^4}}{\displaystyle{(p^2+M_\pi^2)^2}}\,\, 
F_{\pi NN}^2(p^2)}} . \label{ratio_K_eta2} 
\end{eqnarray}
The energy denominators in Eqs. (\ref{ratio_K_eta1}) and (\ref{ratio_K_eta2}) 
ensure that the contributions of the kaon and $\eta$-meson cloud to the 
$\pi N$ sigma-term are negligible when compared to the pionic one. The same 
conclusion regarding the suppression of $K$ and $\eta$-meson loops was 
obtained in the cloudy bag model \cite{Stuckey}. Numerically, kaon and 
$\eta$-meson cloud contributions are $\sigma_{\pi N}^{K} = 1.7 \pm 0.4$ MeV  
and $\sigma_{\pi N}^{\eta} = 0.023 \pm 0.006$ MeV with 
$r_K = 5.6 \pm 1.6 \, \%$  and $r_\eta = 0.08 \pm 0.02 \, \%$.  
 
For the $\pi N$ sigma-term we have the following final value: 
\begin{eqnarray}
\sigma_{\pi N} = \sum\limits_{\Phi=\pi, K, \eta} \sigma_{\pi N}^\Phi = 
45 \pm 5 \,  {\rm MeV}. 
\end{eqnarray}
Our result for the $\pi N$ sigma-term is in perfect agreement with the 
value of $\sigma_{\pi N} \simeq 45$ MeV deduced by Gasser, Leutwyler and 
Sainio \cite{GLS} using dispersion-relation techniques and exploiting the 
chiral symmetry constraints. 

In our opinion, a meaningful description of the $\pi N$ sigma-term should be  
based on the following guide lines: chiral symmetry constraints, fulfilment 
of low-energy theorems, consistency with the Feynman-Hellmann theorem and 
proper treatment of sea-quarks, that is meson cloud contributions. 
The model developed here fulfils these aspects. In the literature there has  
been considerable effort to determine the $\pi N$ sigma-term in the framework 
of different approaches: effective field theories, lattice QCD, QCD sum rules, 
quark models, soliton-type approaches, etc. 
In HBChPT \cite{VKM}-\cite{Borasoy} the $\pi N$ sigma-term is used as an 
input parameter to fix the couplings in the effective Lagrangian. Then the 
$KN$ sigma-terms are predicted (see discussion later on). A detailed analysis 
of the $\pi N$ sigma-term was done in the cloudy bag model 
(CBM) \cite{Stuckey,Jameson},  which is similar to our approach. Numerical 
results obtained in the CBM \cite{Stuckey} and the PCQM are quite close. For 
example, in both calculations the dominant contribution to the $\pi N$ 
sigma-term is due to the pion cloud: $22.7$ MeV (in CBM) and $\approx 30 $ 
MeV (in PCQM). Contributions of kaon and $\eta$-meson loops are significantly 
suppressed in both approaches. A determination of $\sigma_{\pi N}$ in lattice 
QCD has been undertaken by several groups using different techniques: direct 
calculation of the scalar density matrix element of the nucleon and use of the 
Feynman-Hellmann theorem (see detailed discussion in Ref. \cite{Leinweber}). 
As correctly pointed out in Ref. \cite{Leinweber}, the main disadvantage of 
all lattice approaches based on the calculation of the scalar matrix element 
of the nucleon is that latter quantity and the current quark mass $\hat m$ 
are not renormalization group (RG) invariant. This leads to uncertainties in 
the evaluation of the $\pi N$ sigma-term which should be RG invariant. In a 
recent paper \cite{Leinweber} (see also Ref. \cite{Leinweber1}) the $\pi N$ 
sigma-term was calculated by means of the Feynman-Hellmann theorem and using 
present SU(2) lattice QCD data for the nucleon mass as a function of the 
current quark mass. The main idea of Ref. \cite{Leinweber} is to include the 
meson cloud contributions properly as based on chiral symmetry constraints; 
the corresponding expression for the nucleon mass as function of $\hat{m}$ 
includes meson loop contribution ($\pi N$ and $\pi\Delta$) with correct 
leading and next-to-leading order nonanalytic behaviour \cite{Leinweber1}. 
Applying this extrapolation function to the $SU(2)$ lattice data results 
in a value of $\sigma_{\pi N} = 45 \div 55$ MeV. This result is quite close 
to our estimate of the $\pi N$ sigma-term in the two-flavor case 
(see Eq. (\ref{Sigma_piN_pi})). Another example for the dominance of the pion 
cloud contributions to the $\pi N$ sigma-term are the soliton-type quark 
models: the chiral quark soliton model \cite{Diakonov} and the confining 
chiral soliton model \cite{Birse}. Their results  $\sigma_{\pi N} = 54.3$  
MeV \cite{Diakonov} and $\sigma_{\pi N}= 30 \div 40 $ MeV  \cite{Birse} are 
also in qualitative agreement with the previously stated approaches (cloudy 
bag model \cite{Stuckey}, lattice QCD \cite{Leinweber}) and the result 
obtained here. 

One of the advantages of our approach is the possibility to estimate free 
coupling constants in the effective Lagrangians of ChPT, Baryon ChPT (BChPT) 
and HBChPT. For example, in BChPT \cite{Leutwyler1} and 
HBChPT \cite{Meissner1} the quadratic term $(-4c_1M_\pi^2)$ in the chiral 
expansion of the $\pi N$ sigma-term contains the unknown coupling $c_1$ of the 
effective ChPT Lagrangian \cite{Leutwyler1}. Using Eq. (\ref{c2_c4}) we 
predict $c_1=-k_2/4=-1.16\pm 0.1 \, {\rm GeV}^{-1}$. 
This prediction is in good agreement with the value of 
$c_1=-0.925$ GeV$^{-1}$ determined by Becher and Leutwyler  \cite{Becher} 
from a fit of the elastic $\pi N$ scattering amplitude at threshold to 
data of the Karlsruhe partial wave analysis (KA86 data). 

Next we discuss our prediction for the strangeness content of the nucleon 
$y_N$ which is defined in the PCQM as 
\begin{eqnarray}
y_N = \frac{2 <p|S_s^{PCQM}|p>}{ <p|S_u^{PCQM} \, + \, S_d^{PCQM}|p>} . 
\end{eqnarray}
The direct calculation of the strange-quark scalar density $<p|S_s^{PCQM}|p>$ 
using Eq. (\ref{Key_Eq}) is completely consistent with the indirect one 
applying the Feynman-Hellmann theorem 
$<p|S_s^{PCQM}|p> = \partial m_N/\partial m_s$ with 
\begin{eqnarray}\label{bar s s}
<p|S_s^{PCQM}|p> &=& \frac{\partial m_N}{\partial m_s} = 
B \biggl[ \frac{\partial}{\partial M_K^2} + 
\frac{4}{3} \frac{\partial}{\partial M_\eta^2} \biggr] \, m_N \\
&=&\biggl(\frac{g_A}{\pi F}\biggr)^2 B 
\int\limits_0^\infty dp\, p^4 F_{\pi NN}^2(p^2)
\biggl[\frac{d_N^K}{(p^2+M_K^2)^2} \, + \, 
       \frac{4}{3} \, \frac{d_N^\eta}{(p^2+M_\eta^2)^2}\biggr] . 
\nonumber 
\end{eqnarray}
Combining Eqs. (\ref{Sigma_piN}) and (\ref{bar s s}) we derive for $y_N$ 
following expression  
\begin{eqnarray}\label{y_N_PCQM}
\hspace*{-.25cm}
y_N = \frac{
\displaystyle{
\biggl(\frac{g_A}{\pi F}\biggr)^2 B 
\int\limits_0^\infty dp\, p^4 F_{\pi NN}^2(p^2) 
\biggl[ \frac{\displaystyle{d_K}}{\displaystyle{(p^2+M_K^2)^2}} \, + 
\frac{4}{3} \, 
\frac{\displaystyle{d_\eta}}{\displaystyle{(p^2+M_\eta^2)^2}}\biggr]}}
{\displaystyle{\frac{3}{2} \gamma 
\, + \, \biggl(\frac{g_A}{\pi F}\biggr)^2 B 
\int\limits_0^\infty dp\, p^4 F_{\pi NN}^2(p^2) 
\biggl[\frac{\displaystyle{d_\pi}}{\displaystyle{(p^2+M_\pi^2)^2}} \, + 
\frac{1}{2} \, \frac{\displaystyle{d_K}}{\displaystyle{(p^2+M_K^2)^2}} \, + 
\frac{1}{3} \, \frac{\displaystyle{d_\eta}}{\displaystyle{(p^2+M_\eta^2)^2}}
\biggr]}} , 
\end{eqnarray}
with the numerical value of 
\begin{eqnarray}
y_N = 0.076 \pm 0.012 . 
\end{eqnarray}
The small value of $y_N$ in our model is due to the suppressed contributions 
of kaon and $\eta$-meson clouds. Our prediction for $y_N$ is smaller than the 
value $y_N \simeq 0.2$ obtained in \cite{GLS} from an analysis of 
experimental data on $\pi N$ phase shifts. On the other hand, our prediction 
is quite close to the result obtained in the Skyrme model 
$y_N \approx 0.058$ \cite{Blaizot} and in  the cloudy bag model 
$y_N \approx 0.05$ \cite{Stuckey}.  
A revisited prediction of HBChPT for $y_N$ gives $y_N=0.25 \pm 0.05$  
(without inclusion of the decuplet) and  $y_N=0.20 \pm 0.12$ 
(taking into account the decuplet contribution) \cite{Borasoy}. 
Preliminary analyses of the strange content of the nucleon in lattice 
approaches gives larger values for $y_N=0.36\pm 0.03$ \cite{Dong} 
and $y_N=0.59\pm 0.13$ \cite{Gusken}, which imply a big contribution of the 
strange quark sea to the nucleon mass. 

Next, we present our results for the $KN$ sigma-terms 
\begin{eqnarray}
& &\sigma_{KN}^u \equiv  \sigma_{KN}^{(1)}= 340 \pm 37 \, {\rm  MeV},  
\hspace*{.5cm} \sigma_{KN}^d  =  284 \pm 37 \, {\rm MeV}, \\
& &\sigma_{KN}^{(2)} = 228 \pm 37 \, {\rm MeV}, \hspace*{1.85cm}
\sigma_{KN}^{I=1} = 28 \, {\rm MeV} \nonumber\\ 
\mbox{and} \,\,\,\,\, & &\sigma_{KN} \equiv 
\frac{\sigma_{KN}^u+\sigma_{KN}^d}{2} = 312 \pm 37 \, {\rm MeV}. 
\nonumber
\end{eqnarray}
Comparative results for the $KN$ 
sigma-term are less abundant than for the $\pi N$ sigma-terms. 
We cite recent (revisited) results of HBChPT (taking into account the decuplet 
contribution)  with $\sigma_{KN}^{(1)}= 380 \pm 40 $  MeV,  
$\sigma_{KN}^{(2)}= 250 \pm 30 $ MeV \cite{Borasoy}, results of lattice QCD 
with $\sigma_{KN}=(\sigma_{KN}^u+\sigma_{KN}^d)/2=362\pm 13$ MeV \cite{Dong} 
and predictions of the Nambu-Jona-Lasinio model with $\sigma_{KN}=425$ MeV 
(with an error bar of 10-15\%) \cite{Hatsuda_Kunihiro}. 
Hopefully, future DA$\Phi$NE experiments at Frascati \cite{Gensini} 
will allow for a determination of the $KN$ sigma-terms and hence for a better 
knowledge of the strangeness content of the nucleon. 
 
Finally, for the sake of completeness we present the set of our predictions 
for the $\Delta$-isobar condensates and $\eta N$ sigma-term: 
\begin{eqnarray}
& &\sigma_{\pi\Delta}=32 \pm 3 \, {\rm MeV}, \hspace*{1cm} 
y_\Delta=0.12 \pm 0.02, \hspace*{1.3cm} 
\sigma_{\eta N}=72 \pm 16 \, {\rm MeV}, 
\\
& &\sigma_{K\Delta}^u=262 \pm 26 \, {\rm MeV}, \hspace*{.5cm} 
\sigma_{K\Delta}^d=206 \pm 26 \, {\rm MeV},   \hspace*{.5cm} 
\sigma_{\eta\Delta}=75 \pm 17 \, {\rm MeV}.
\nonumber
\end{eqnarray}

\section{Summary and conclusions} 

In conclusion, we have evaluated the meson-baryon sigma-terms using a 
perturbative chiral quark model based on an effective chiral Lagrangian. 
The Lagrangian describes baryons as bound states of three valence 
quarks surrounded by a cloud of pseudoscalar mesons as dictated by 
the chiral symmetry requirement. The calculated quantities contain only one 
model parameter $R$, which is related to the radius of the three-quark core,  
and are otherwise expressed in terms of 
fundamental parameters of low-energy hadron physics: axial coupling constant 
$g_A$, weak pion decay constant $F$, normalized strong pion-nucleon form 
factor $F_{\pi NN}$ and set of QCD parameters (current quark 
masses $\hat m$ and $m_s$ and quark condensate parameter $B$). 
Predictions are given for a variation of the free parameter $R$ in a quite 
wide physical region from 0.55 Fm to 0.65 Fm corresponding to 
$<r^2_E>^P_{3q-core}$ ranging from 0.5 Fm$^2$ to 0.7 Fm$^2$. Our result for 
the $\pi N$ sigma-term is in a perfect agreement with the value obtained 
by Gasser, Leutwyler and Sainio \cite{GLS} and with the results of other 
theoretical approaches (cloudy bag model \cite{Stuckey}, chiral quark 
soliton model \cite{Diakonov}, lattice QCD \cite{Leinweber}, etc. ) where 
the pion cloud contribution properly was taken into account. Analyses of the 
strangeness content of the nucleon, kaon-baryon and eta-baryon sigma-terms 
were done in detail. We compare our predictions to results of other 
theoretical approaches. Ongoing experimental efforts aim to allow for a 
reliable extraction of these quantities, and hence to fix the scalar 
strange-quark density in the nucleon. To solidify the model 
approach we mention that with the same values of the free model parameter $R$  
we recently obtained \cite{PCQM} a quite reasonable description of static 
properties and electromagnetic form factors of the nucleon. 

{\it Acknowledgements}. This work was supported by the Deutsche 
Forschungsgemeinschaft (DFG, grant FA67/25-1 and 436 RUS 113/595/0-1). 
One of us (E.G.D.) acknowledges the support of RFBR by the grant 00-02-16853.

\begin{figure}[t]
\noindent FIG.1: Diagrams contributing to the baryon energy shift:  

\noindent meson cloud (1a) and meson exchange diagram (1b).

\vspace*{1cm}
\noindent FIG.2: Diagrams contributing to the meson-baryon sigma-terms: 

\noindent tree diagram (2a), meson cloud diagram (2b) and  
meson exchange diagram (2c).
 
\noindent Insertion of the scalar density operator is depicted 
by the symbol $" \bf \vee "$. 

\end{figure}

\begin{figure}
\centering{\
\epsfig{figure=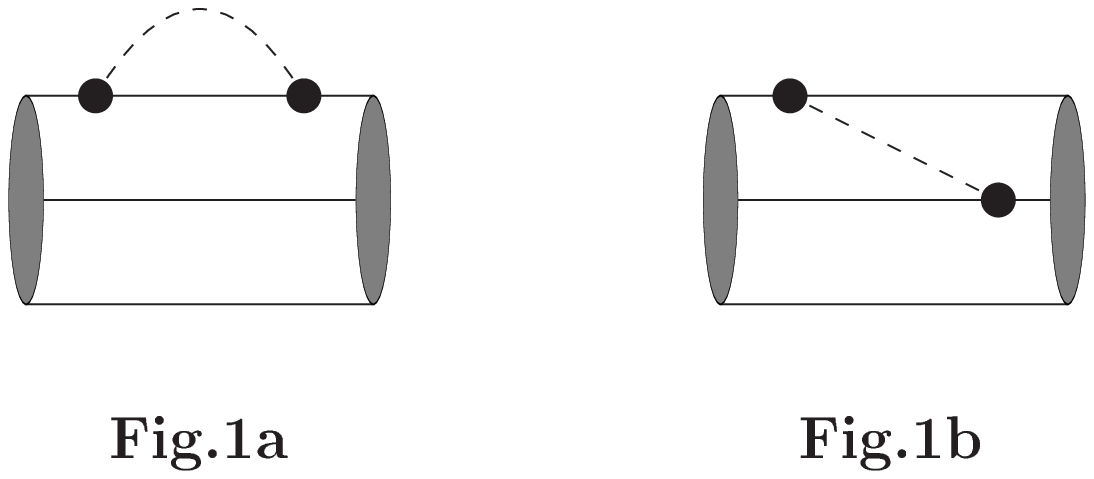,height=21cm}}
\end{figure}

\begin{figure}
\centering{\
\epsfig{figure=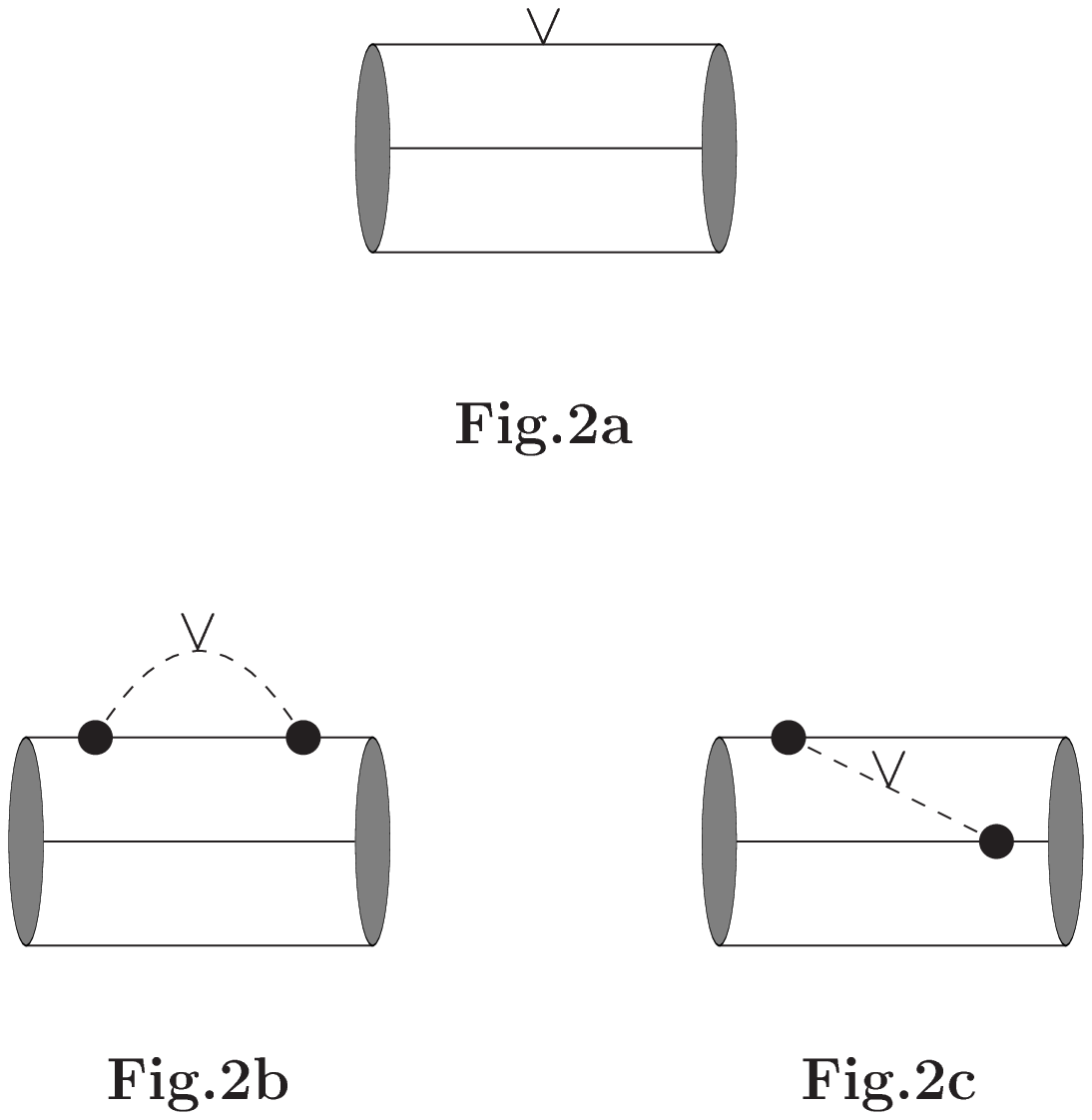,height=21cm}}
\end{figure}

\end{document}